\documentclass[preprint,prd,showpacs]{revtex4} 
 
\usepackage{amsmath}
\usepackage{amsfonts}
\begin{document} 

\title{\bf Possible Constraints on the Duration of Inflationary Expansion from
Quantum Stress Tensor Fluctuations}
\author{Chun-Hsien Wu}
\email{chunwu@phys.sinica.edu.tw} 
\author{Kin-Wang Ng}
\email{nkw@phys.sinica.edu.tw}
\affiliation{Institute of Physics \\
Academia Sinica \\
Nankang, Taipei 11529 \\
 Taiwan}
\author{L.H. Ford}
\email{ford@cosmos.phy.tufts.edu}
\affiliation{Institute of Cosmology \\ 
Department of Physics and Astronomy \\
Tufts University, Medford, MA 02155}
\date{\today}

\begin{abstract}
We discuss the effect of quantum stress tensor fluctuations in deSitter 
spacetime upon the expansion of a congruence of timelike geodesics.
We treat a model in which the expansion fluctuations begin on a given 
hypersurface in deSitter spacetime, and
 find that this effect tends to grow, in contrast to the situation
in flat spacetime. This growth potentially leads to observable consequences
in inflationary cosmology in the form of density perturbations which
depend upon the duration of the inflationary period. In the context of our
model, the effect may be used to place upper bounds on this duration. 
\end{abstract}


\maketitle

\baselineskip=14pt 

\section{Introduction}
\label{sec:intro}

Quantum fluctuations of the stress tensor operator have been studied
in numerous recent papers
\cite{F82,KF93,HS98,H99,PH01,HV03,CH95,CCV,RV,Stochastic,Moffat,Borgman,BF04b,WF99,WF01,FW03,FW04,FR05,Wu06}. Fluctuations of the stress tensor drive
passive fluctuations of the gravitational field, which are to be distinguished
 from the active fluctuations due to the quantization of the gravitational
degrees of freedom. Stress tensor fluctuations play a crucial role
in stochastic gravity, and their role in the early universe has been
discussed by several authors.

In the present paper, we will be concerned with the effects of quantum
stress tensor fluctuations in inflationary cosmology. Since the pioneering
paper by Guth~\cite{Guth}, the inflationary paradigm has been extensively 
developed and now seems to be in good agreement with 
observations~\cite{Liddle,Spergel}. One of the successes of inflation is a
natural solution of the horizon problem; inflationary expansion allows
the entire observable universe today to have arisen from a region within which 
all parts were once in causal contact. The rapid expansion smooths
any initial classical perturbations, and leads to a subsequent universe
which is relatively independent of the duration of inflation, so long as
there is inflation by at least a factor of $10^{23}$. However, it can
be shown that inflation could not have had an infinite duration in the 
past~\cite{BGV03}. 
A key prediction
of inflationary cosmology is a nearly flat spectrum of initial density
perturbations, which arise from the intrinsic quantum fluctuations of
an inflaton field. However, quantum stress tensor fluctuations of all
quantum fields, not just the inflaton, should also contribute to density
perturbations by means of passive metric fluctuations. Unlike the
fluctuations of a nearly free inflaton field, quantum stress tensor 
fluctuations are expected to have a non-Gaussian probability distribution.
Evidence for this comes from calculations in simple cases which reveal
that in general the third moment is nonzero, and hence the probability 
distribution cannot be symmetric~\cite{F06,FFR}. 

Here we will consider the effects of stress tensor fluctuations of
the conformally invariant scalar field and the electromagnetic field
in deSitter spacetime upon geodesics of test particles. We will employ
the Raychaudhuri equation to calculate fluctuations in the expansion
$\theta$ of a congruence of comoving timelike geodesics, and then use the
results to draw inferences about density perturbations in the post-inflationary
period. The outline of this paper is as follows: Some basic formalism
will be developed in Sec.~\ref{sec:formalism}. This formalism will be applied
to inflationary cosmology in Sec.~\ref{sec:First}.  We show that
$\theta$-fluctuations build up during inflation and influence the redshifting
of radiation after inflation. In this model, we will make no specific
references to the mechanism by which inflation ends, but will show
that the resulting  density perturbations grow as the length of the 
inflationary epoch increases. This will lead to an upper bound on the 
duration of inflation. In Sec.~\ref{sec:Second}, we will examine the
effects of $\theta$-fluctuations on the dynamics of an inflaton field,
and show that there is a further mechanism by which stress tensor 
fluctuations create density perturbations  sensitive to the length of
inflation. This model leads to a stronger bound on the duration of inflation.
The results will be discussed in Sec.~\ref{sec:Final}.

\section{Basic Formalism}
\label{sec:formalism}

\subsection{The Raychaudhuri equation and the conservation law}

The key tool which we will employ for studying the effects of passive
metric fluctuations in deSitter spacetime will be the Raychaudhuri equation,
which can be written for a congruence of timelike geodesics with 
four-velocity $u^\mu$ as
\begin{equation}
\frac{d \theta}{d \lambda} = - R_{\mu\nu} u^\mu u^\nu - \frac{1}{3}\, \theta^2
-\sigma_{\mu\nu} \sigma^{\mu\nu} + \omega_{\mu\nu}  \omega^{\mu\nu} \,.
                                                 \label{eq:ray}
\end{equation}
Here $\theta=u^\mu_{;\mu}$ is 
the expansion of the congruence and $\lambda$ is an
affine parameter. In addition, $R_{\mu\nu}$ is the Ricci tensor, 
$\sigma^{\mu\nu}$ is the shear, and $\omega^{\mu\nu}$ is the vorticity
of the congruence. The vorticity may be set to be zero, and we will assume that
the shear is negligible. In this case, the equation reduces to
\begin{equation}
\frac{d \theta}{d \lambda} = - R_{\mu\nu} u^\mu u^\nu - \frac{1}{3}\, \theta^2
 \,.   \label{eq:ray2}
\end{equation}
Consider the case of a Robertson-Walker spacetime, where the metric
can be written in terms of the scale factor $a(t)$ as
\begin{equation}
ds^2 = -dt^2 +a^2(t) (dx^2+dy^2+dz^2) \,.
\end{equation}
In the case of comoving geodesics which remain at rest
in these coordinates, $u^\mu = \delta_t^\mu$ and the expansion is given by
\begin{equation}
\theta = \theta_{0}=3\frac{\dot{a}}{a}\,,
   \label{eq:theta0}
\end{equation}
where $\dot{a} = da/dt$.
With a perfect fluid source, the matter stress tensor is
\begin{equation}
T_{\mu\nu}=(p+\rho)\, u_{\mu}u_{\nu}+p\, g_{\mu\nu}\,,
  \label{eq:fluid}
\end{equation}
where $\rho$ is the energy density and $p$ is the pressure. 
The Ricci tensor is given in terms of the stress tensor by Einstein's
equations
\begin{equation}
R_{\mu\nu}  =  8\pi\left(T_{\mu\nu}-\frac{1}{2}\, g_{\mu\nu}T\right)\,,
  \label{eq:Ricci}
\end{equation}
and for the perfect fluid we have
\begin{equation}
R_{\mu\nu}u^{\mu}u^{\nu}=4\pi(\rho+3p)\,.
   \label{eq:classical_Ricci}
\end{equation}
Thus $\theta_0$ satisfies
\begin{equation}
\frac{d\theta_{0}}{dt}= -4\pi(\rho+3p) -\frac{1}{3}\theta_{0}^{2}\,.
\end{equation}

The conservation law for a perfect fluid can be expressed in terms of
the expansion as~\cite{Hawking66,Olson}
\begin{equation}
\dot{\rho}+(\rho+p)\,\theta =0 \,.
  \label{eq:conservation}
\end{equation}
Write the equation of state as $p=w\,\rho$ and assume $w$ is constant.
The density can be expressed as
\begin{equation}
\rho(t)=\rho(0)\, e^{-(1+w)\int_{0}^{t}\theta(t_{1})\, dt_{1}}\,.
   \label{eq:rho}
\end{equation}
For the case of unperturbed Robertson-Walker spacetime, this is equivalent
to
\begin{equation}
\rho(t)=\rho_{0}(t)=\rho(0)\,\left[\frac{a(t)}{a(0)}\right]^{-3(1+w)}\,.
\label{eq:rho2}
\end{equation}

\subsection{Fluctuations of the expansion}

Now we wish to consider perturbations of the background spacetime
produced by stress tensor fluctuations. Let $\theta =\theta_0 +\theta_1$,
$\rho = \rho_0 + \delta \rho$, and $p = p_0 + \delta p$. For now, we
treat $\delta \rho$, and $\delta p$ as independent variables.
Let the Ricci tensor term in the Raychaudhuri equation be expressed as a
sum of a classical part, given by Eq.~(\ref{eq:classical_Ricci}),
and a smaller fluctuating quantum part, denoted by 
$(R_{\mu\nu}u^{\mu}u^{\nu})_{q}$. Note that the quantum field responsible
for the fluctuations is distinct from the classical perfect fluid.
Because of the possibility of pressure gradients, the fluid elements will not
in general move along geodesics and the Raychaudhuri equation acquires an
additional term on the right hand side~\cite{LS90,LL93} of
\begin{equation}
- \frac{\nabla^2 \delta p}{\rho + p}\,,
\end{equation}
where $\nabla^2$ is the Laplacian operator in a constant $t$ hypersurface.
If we expand the  Raychaudhuri equation
to first order in $\theta_1$, $\delta\rho$, $\delta p$ and 
$(R_{\mu\nu}u^{\mu}u^{\nu})_{q}$, we find
\begin{equation}
\frac{d\theta_{1}}{dt}=-4\pi(\delta\rho + 3 \delta p) 
- \frac{\nabla^2 \delta p}{\rho_0 + p_0}
- \left(R_{\mu\nu}u^{\mu}u^{\nu}\right)_{q}-
\frac{2}{3}\theta_{0}\theta_{1}\,.
\label{eq:theta1}
\end{equation}
This equation may be integrated to find
\begin{equation}
\theta_{1}(t)=-a^{-2}(t)\int_{t_{0}}^{t}dt'\, a^{2}(t')
\left[4\pi(\delta\rho(t')+ 3 \delta p(t')) 
- \frac{\nabla^2 \delta p}{\rho_0 + p_0}
-\left(R_{\mu\nu}u^{\mu}u^{\nu}\right)_{q}\right]\,,
\label{eq:theta1b}
\end{equation}
with the initial condition $\theta_1(t_0) =0$.

To leading order, we may regard $\delta\rho$ and $\delta p$ as 
perturbations which have some source other than quantum stress tensor
fluctuations. In most inflationary models, these perturbations are driven
by quantum fluctuations of the inflaton field, but for our purposes they
can be treated as being either classical, or at least uncorrelated with
the stress tensor fluctuations. Thus it is convenient to split $\theta_{1}$
into a ``classical'' part $\theta_{1c}$ which depends upon $\delta\rho$ and 
$\delta p$, and a quantum part, $\theta_{1q}$, driven by the stress tensor
fluctuations. Then $\theta_{1} = \theta_{1c} + \theta_{1q}$, and
\begin{equation}
\theta_{1q}(t)=-a^{-2}(t)\int_{t_{0}}^{t}dt'\, a^{2}(t')
\left(R_{\mu\nu}u^{\mu}u^{\nu}\right)_{q}\,.
\label{eq:theta1q}
\end{equation}
Thus $\theta_{1q}$ is given by the fluctuating part of the Ricci tensor, which
is in turn given by the quantum stress tensor through Eq.~(\ref{eq:Ricci}).
In this case, we can construct a correlation function for the expansion
as a integral of the Ricci tensor correlation function:
\begin{equation}
K_{\mu\nu\alpha\beta} = \langle  R_{\mu\nu}(x)\, R_{\alpha\beta}(x') \rangle
- \langle  R_{\mu\nu}(x) \rangle \langle R_{\alpha\beta}(x') \rangle \,,
\end{equation}
and write
\begin{equation}
\langle \theta(t_1)\, \theta(t_2) \rangle - 
\langle \theta(t_1)\rangle \langle  \theta(t_2) \rangle =
a^{-2}(t_1)\,a^{-2}(t_2)\, \int_{t_{0}}^{t_{1}}dt\, a^{2}(t)
\int_{t_{0}}^{t_{2}}dt'\, a^{2}(t')\; 
K_{\mu\nu\alpha\beta}\,u^{\mu} u^{\nu} u^\alpha u^\beta\,.
   \label{eq:theta_corr}
\end{equation}
Note that if $\theta_{1c}$ is truly classical and non-fluctuating,  
we do not need to make a distinction between $\theta$ and
$\theta_{1q}$ in the above expression, as the correlation function for 
both is the same: only 
the fluctuating part contributes to the correlation function.
If $\delta\rho$ and $\delta p$ do indeed fluctuate, then we assume that
their fluctuations are uncorrelated with those of the stress tensor, so their
effect would be to add another term in the expansion correlation function.
This assumption will be discussed in more detail later.
In writing Eq.~(\ref{eq:theta_corr}), we are essentially assuming that the 
$\theta$-fluctuations vanish before the $t=t_0$ hypersurface, which amounts
to a sudden switching assumption.

In this paper, we will restrict our attention to passive metric
fluctuations caused by conformally invariant quantum fields. In this case,
the classical stress tensor in Robertson-Walker spacetime is related
to that in flat spacetime by a conformal transformation:
\begin{equation}
T_{\mu\nu}^{RW}(x) = a^{-4}(t)\, T_{\mu\nu}^{flat}(x) \,.
\end{equation}
The quantum stress tensor operator in curved spacetime has an anomalous
trace, so curved spacetime expectation values cannot be obtained directly
by a conformal transformation of the corresponding flat spacetime
expectation value. However, the contribution of the anomalous trace to
the stress tensor operator is a c-number, and hence will cancel in a
stress tensor correlation function~\cite{birrell}. 
Thus we can express the correlation 
function for the stress tensor in Robertson-Walker spacetime as the
conformal transform of the corresponding flat space correlation function,
\begin{equation}
C_{\mu\nu\alpha\beta}^{RW}(x,x') = a^{-4}(t)\,a^{-4}(t')\,
C_{\mu\nu\alpha\beta}^{flat}(x,x')\, ,
\end{equation}
where
\begin{equation}
C_{\mu\nu\alpha\beta}(x,x') = 
\langle  T_{\mu\nu}(x)\, T_{\alpha\beta}(x') \rangle
- \langle  T_{\mu\nu}(x) \rangle \langle T_{\alpha\beta}(x') \rangle \,.
\end{equation}
Because the anomalous trace of the stress tensor does not contribute to 
correlation functions, we can use the Einstein equation, Eq.~(\ref{eq:Ricci}),
to relate the Ricci and stress tensor correlation functions:
\begin{equation}
K_{\mu\nu\alpha\beta}(x,x') = (8 \pi)^2\, C_{\mu\nu\alpha\beta}(x,x')\,.
\end{equation}

Here we will assume that the quantum field in Robertson-Walker spacetime
is in the conformal vacuum state, so the corresponding flat space
correlation function will be that for the Minkowski vacuum state.  
In this case, we have
\begin{equation}
K_{\mu\nu\alpha\beta}(x,x')u^{\mu}u^{\nu}u^{\alpha}u^{\beta} =
(8 \pi)^2\, C_{ttt't'}(x,x')= (8 \pi)^2\,a^{-4}(t)\,a^{-4}(t')\,{\cal E}\,,
\end{equation}
where ${\cal E}$ is the flat space vacuum energy density correlation
function.

\subsection{Treatment of singular integrands}
\label{sec:singular}

Observable quantities are expressed as integrals of stress tensor correlation 
functions, as in Eq.~(\ref{eq:theta_corr}). However, the integrands in
these integrals contain singularities which appear not to be integrable,
specifically higher order poles on the real axis. Nonetheless, such integrals
can be given an unambiguous, finite value. One approach which could be
used is dimensional regularization. This approach has been studied in flat
spacetime in Ref.~\cite{FW04}, where it was shown that 
spacetime integrals of stress tensor 
correlation functions such as $C_{\mu\nu\alpha\beta}(x,x')$ are 
actually finite in dimensional regularization. 
This means that if we were to evaluate an integral of
$C_{\mu\nu\alpha\beta}(x,x')$ in spacetime dimension $4+ \varepsilon$, and 
then take the limit that $\varepsilon \rightarrow 0$, the result will be
finite. (This is not true for integrals of 
correlation functions involving a time-ordered
product of stress tensor operators. In this case, the singularity as
$\varepsilon \rightarrow 0$ is proportional to counterterms in the 
gravitational action quadratic in the curvature.) 

An alternative approach, which is easier to use in practice, is to define
the integrals by an integration by parts procedure. This is the generalized
principal value discussed in Ref.~\cite{Davies}, and used in 
Refs.~\cite{Borgman,WF01,FW03,FR05}. The basic idea is to re-express an 
integral of the form 
\begin{equation}
\int_a^b \frac{f(x)}{(x-c)^n} \, dx
\end{equation}
by use of the identity       
\begin{equation}
 \frac{1}{(x-c)^n} = 
(-1)^{n-1} (n-1)!\, \frac{d^{n}}{dx^{n}}\, \ln(x-c)  \, ,
\end{equation}
and then perform successive integrations by parts to express the original
integral as a sum of finite boundary terms and an integral whose integrand
has only a logarithmic singularity. An equivalent approach is to seek an
antiderivative, $G(x)$ of the function $F(x) = f(x)/(x-c)^n$, that is,
$G'(x) = F(x)$, and write
\begin{equation}
\int_a^b F(x)\, dx = G(b) - G(a) \,.
\end{equation}
This result would be trivial if $F(x)$ had no singularities. In the present
case, the only nontrivial aspect arises from whether the contour of 
integration goes above or below the higher order pole at $x=c$. However,
the residue of this pole is pure imaginary if $f(x)$ and its derivatives are 
real. This is the case in the present problem, where one needs to integrate
a real valued correlation function to get a real answer. Consequently, 
the residue of the pole will not contribute to the final result.

\section{First Model: $\theta$-Fluctuations and Redshifts after Reheating}
\label{sec:First}

In this section, we will examine the fluctuations of the expansion during
an inflationary period, and their subsequent effects in creating density
fluctuations by differential redshifts after the end of inflation.
It will be convenient to use the conformal time $\eta$ rather than the 
proper time $t$ of the comoving observers. The two time coordinates are 
related by
\begin{equation}
d \eta = a^{-1}(t) dt \,.
\end{equation}
Let the inflationary phase begin at $\eta = \eta_0$ and end at $\eta =0$.
During this interval, the spacetime will be taken to be deSitter space, 
which may be represented as a spatially flat Robertson-Walker metric
with
\begin{equation}
a(\eta) =  \frac{1}{1 - H \eta}\,,  \qquad \eta_0 \leq \eta \leq 0 \,.
\end{equation}
For $\eta \geq 0$, we take the scale factor to be that of a radiation-dominated
universe, for which
\begin{equation}
a(\eta) = 1+ H\, \eta \,.
 \label{eq:a_rad1}
\end{equation}
These forms are chosen so that both $a(\eta)$ and its first derivative
are continuous at $\eta = 0$.
 In terms of comoving time, the scale factors are
\begin{equation}
a(t) = {\rm e}^{H (t-t_R)}\,, \quad t \leq t_R\,,  \label{eq:a_deS} 
\end{equation}
and 
\begin{equation}
a(t) = \sqrt{1 + 2H\,(t-t_R)}\,, \quad t \geq t_R\,,
   \label{eq:a_rad2}
\end{equation}
where $t=t_R$ is the comoving time at which inflation ends.

The expansion correlation function, both during and after inflation,
may be written as
\begin{equation}
\langle \theta(\eta_1)\, \theta(\eta_2) \rangle - 
\langle \theta(\eta_1)\rangle \langle  \theta(\eta_2) \rangle =
(8\,\pi)^2\; a^{-2}(\eta_1)\,a^{-2}(\eta_2)\, \int_{\eta_{0}}^{\eta_{1}}
\frac{d\eta}{a(\eta)}
\int_{\eta_{0}}^{\eta_{2}}
\frac{d\eta'}{a(\eta')}\; {\cal E}(\Delta\eta,r) \,,
   \label{eq:theta_corr2}
\end{equation}
where $\Delta\eta = \eta -\eta'$ and $r = |\mathbf{x} - \mathbf{x'}|$ is the
coordinate space separation of the pair of points at which $\theta$ is
measured. Here we will assume that the $\theta$-fluctuations vanish
at the beginning of inflation, $\eta = \eta_0$.
During inflation, there is no classical matter present. However, after
reheating at $\eta = 0$, variations in $\theta$ cause the matter in
different spatial regions to redshift at different rates, leading to
variations in the density of the classical matter. Once reheating has 
occurred, $\delta p = w\, \delta \rho$, where $w = 1/3$ in our model. 
Equation~(\ref{eq:rho}) for the energy density can be written as 
\begin{eqnarray}
\rho(t) & = & \rho(0)\, e^{-(1+w)\int_{0}^{t} \theta_0 \, dt_{1}}
\; e^{-(1+w)\int_{0}^{t}\theta_{1}(t_{1})\, dt_{1}}\nonumber \\
 & \approx & \rho_{0}(t)\,\left[1
-(1+w)\int_{0}^{t}\theta_{1}(t_{1})\, dt_{1}+O(\theta_{1}^{2})\right]\,,
 \label{eq:rho3}
\end{eqnarray}
so that
\begin{equation}
\frac{\delta\rho}{\rho_{0}}=
-(1+w)\int_{t_R}^{t}\theta_{1}(t_{1})\, dt_{1}
=  -(1+w)\int_0^\eta \frac{d\eta_1}{a(\eta_1)}\,\theta_{1}   \,.
\label{eq:delrho}
\end{equation}
The expansion fluctuations lead to density fluctuations given by
\begin{equation}
\left\langle\left(\frac{\delta\rho}{\rho}\right)^{2}\right\rangle=
(8\,\pi)^2\; (1+w)^{2}\int_0^{\eta_{s}}\frac{d\eta_1}{a(\eta_1)}
\int_0^{\eta_{s}} \frac{d\eta_2}{a(\eta_2)} \,
\int_{\eta_{0}}^{\eta_{1}}\frac{d\eta}{a(\eta)}
\int_{\eta_{0}}^{\eta_{2}}\frac{d\eta'}{a(\eta')}\,{\cal E}(\Delta\eta,r)\,.
\label{eq:del_rho}
\end{equation} 
In the above expression, the integrals on $\eta_1$ and $\eta_2$ represent
the differential redshifting, and hence have a lower limit  at $\eta = 0$,
the reheating time. The integrals on $\eta$ and $\eta'$ describe
the effects of quantum stress tensor fluctuations on the expansion
$\theta$. The integration range is the beginning of inflation 
at $\eta = \eta_0$
to a time $\eta =\eta_s$ when the density variations are measured. We can
take $\eta_s$ to be the time of last scattering, when the density
fluctuations of the cosmic background radiation were established.

The expression Eq.~(\ref{eq:del_rho}) contains contributions from all
length scales, whereas observations are sensitive only to a finite
range of scales. Thus, in order to compare the results of our calculations
with observation, we should look at the power spectrum $P_{k}(\eta_{s})$
defined by
\begin{equation}
\left\langle\left(\frac{\delta\rho}{\rho}\right)^{2}\right\rangle  =
\int d^{3}k\, {\rm e}^{i\,\mathbf{k}\cdot\Delta\mathbf{x}}P_{k}(\eta_{s})\,.
\label{eq:power}
\end{equation}
We will first compute the density fluctuations as a function of $r$,
and then Fourier transform the result to obtain $P_{k}(\eta_{s})$.

Let 
\begin{eqnarray}
F(\eta_{1},\eta_{2}) & = & 
\int_{\eta_{0}}^{\eta_{1}}\frac{d\eta}{a(\eta)}
\int_{\eta_{0}}^{\eta_{2}}\frac{d\eta'}{a(\eta')}{\cal E}(\Delta\eta,r)
                                  \nonumber \\
 & = & F_{0}+F_{1}(\eta_{1})+F_{1}(\eta_{2})+F_{2}(\eta_{1},\eta_{2})\, ,
\label{eq:F}
\end{eqnarray}
where
\begin{eqnarray}
F_{0} & = & \int_{\eta_{0}}^{0}\frac{d\eta}{a(\eta)}
\int_{\eta_{0}}^{0}\frac{d\eta'}{a(\eta')}{\cal E}(\Delta\eta,r)\, ,\nonumber \\
F_{1}(\eta_{1}) & = & \int_{0}^{\eta_{1}}\frac{d\eta}{a(\eta)}
\int_{\eta_{0}}^{0}\frac{d\eta'}{a(\eta')}{\cal E}(\Delta\eta,r)\, ,\nonumber \\
F_{1}(\eta_{2}) & = & \int_{\eta_{0}}^{0}\frac{d\eta}{a(\eta)}
\int_{0}^{\eta_{2}}\frac{d\eta'}{a(\eta')}{\cal E}(\Delta\eta,r)\, ,\nonumber \\
F_{2}(\eta_{1},\eta_{2}) & = & \int_{0}^{\eta_{1}}\frac{d\eta}{a(\eta)}
\int_{0}^{\eta_{2}}\frac{d\eta'}{a(\eta')}{\cal E}(\Delta\eta,r)\,.
\label{eq:F_parts}
\end{eqnarray}
Here $F_{0}$ describes correlated stress tensor fluctuations entirely
within the deSitter phase, $F_{2}$ similarly describes fluctuations
entirely in the radiation-dominated phase, and $F_{1}$ describes the 
correlation of fluctuations between the deSitter and radiation-dominated 
phases. Note that
\begin{equation}
F_0 = \langle \theta(0)\, \theta(0) \rangle - 
\langle \theta(0)\rangle \langle  \theta(0) \rangle =
\langle (\Delta \theta)^2 \rangle 
\end{equation}
is the variance of the expansion at the end of inflation.
 
To go further, we need the explicit form for the flat space energy
density correlation function, ${\cal E}(r,\Delta\eta)$. For the case
of the electromagnetic field, it is
\begin{equation}
{\cal E}_{em}= \frac{(r^{2}+3\Delta\eta^{2})^{2}}
{4\pi^{4}(r^{2}-\Delta\eta^{2})^{6}}\,.
\label{eq:EM_corr}
\end{equation}
For the conformal scalar field,
\begin{equation}
{\cal E}_{scalar}=\frac{(\Delta\eta^{2}+3r^{2})(r^{2}+3\Delta\eta^{2})}
{12\pi^{4}(r^{2}-\Delta\eta^{2})^{6}}\,.
\label{eq:scalar_corr}
\end{equation}
Let us first consider $F_{0}$, which may be written as 
\begin{equation}
F_{0}=\int_{0}^{|\eta_{0}|}d\eta\,(1+H\eta)\int_{0}^{|\eta_{0}|}
d\eta'\,(1+H\eta')\,{\cal E}(\Delta\eta,r) \,.
\label{eq:F0}
\end{equation}
This integral may be evaluated using algebraic symbol manipulation programs.
We have used both {\it Mathematica} and {\it Maxima} with equivalent results.
The program is asked to find antiderivatives of the integrand, which are 
then evaluated at the appropriate limits. (See the discussion 
in Sec.~\ref{sec:singular}.) The result is rather complicated, but simplifies
greatly in the limit that $H |\eta_{0}| \gg 1$.
For the electromagnetic field case, we find the asymptotic form
\begin{equation}
F_{0}\approx\frac{8 H^{2}|\eta_{0}|^{2}}{5\pi^{2}r^{6}}
\label{eq:F0_asy}
\end{equation}
for $H |\eta_{0}| \gg 1$. For the scalar case in the same limit,
\begin{equation}
F_{0}\approx\frac{8 H^{2}|\eta_{0}|^{2}}{15\pi^{2}r^{6}} \,.
\label{eq:F0_asy2}
\end{equation} 

The remarkable feature of this result is that it grows with increasing
$|\eta_{0}|$, and hence depends upon the length of the inflationary period.
Note that $F_{2}$ is independent of $|\eta_{0}|$, and $F_{1}$ is found to
go to a finite limit for large $|\eta_{0}|$. Thus in this limit, 
\begin{equation}
F(\eta_{1},\eta_{2}) \approx F_{0}\,.
\end{equation} 
Note that 
\begin{equation}
H |\eta_{0}| \approx  {\rm e}^{H (t_R -t_0)}
\end{equation}
is the net expansion factor during inflation, which needs to be greater
than about $10^{23}$ to solve the horizon problem. Thus the large
$H |\eta_{0}|$ approximation is an extremely good one. Note that there are 
no real particles moving along the comoving geodesics during inflation.
Nonetheless, the expansion $\theta$ at the end of inflation has
observable consequences. If we consider the reheating to occur very quickly,
then we are effectively matching deSitter spacetime and the radiation dominated
Robertson-Walker spacetime across the $\eta = 0$ hypersurface. Any such
matching must satisfy the Israel junction conditions, that the extrinsic
curvature of this hypersurface be continuous. (Note that $\theta$ is 
the trace of the extrinsic curvature tensor of this surface.) 
This implies that $\theta$
must be continuous, even in the case where there are spatial variation
in $\theta$. As a result, the expansion fluctuations generated by stress
tensor fluctuations in the deSitter phase persist in the radiation-dominated
phase and cause density fluctuations. Recall that the geodesics in deSitter 
space with which we are concerned become the comoving geodesics in the
post-inflationary universe. This choice breaks the deSitter invariance.

In order to compute the power spectrum of these fluctuations, 
we must find the Fourier transform
of $1/r^6$, which may be done by integration by parts as follows:
\begin{eqnarray}
& & \frac{1}{(2\pi)^3}\int d^{3}x\,\frac{1}{r^6}\, 
{\rm e}^{-i\,\mathbf{k}\cdot\Delta\mathbf{x}}
= \frac{1}{2\pi^2 k} \int_0^\infty dr \,\frac{1}{r^5}\,\sin(k r)
\nonumber \\
&=& \frac{1}{4\pi^2 k\, 4!} \int_{-\infty}^\infty dr\,\sin(k r)\,
\frac{d^4}{dr^4} \left(\frac{1}{r}\right) 
= \frac{1}{4\pi^2 k\, 4!} \int_{-\infty}^\infty dr\,\left(\frac{1}{r}\right)
\frac{d^4}{dr^4}\,\sin(k r) \nonumber \\
&=& \frac{k^3}{48 \pi^2 } \int_0^\infty dr\,\frac{\sin(k r)}{r}
=  \frac{k^3}{96 \pi} \,.
\end{eqnarray}

For the electromagnetic case, this leads to
\begin{equation}
P_{k}(\eta_{s})\approx
\frac{32 H^{2}|\eta_{0}|^{2} k^{3}}{15\,\pi}
\left(\int_{0}^{\eta_{s}}\frac{d\eta_{1}}{a(\eta_{1})}\right)^{2}(1+w)^{2}\,.
\label{eq:power2}
\end{equation}
We may evaluate the integral in the above expression using 
Eqs.~(\ref{eq:a_rad1}) and (\ref{eq:a_rad2}) to find
\begin{equation}
P_{k}(\eta_{s})\approx
\frac{32 |\eta_{0}|^{2}k^{3}}{15\,\pi}\, 
\ln^2[a(\eta_s)]\,(1+w)^{2}\,\ell_{p}^{4}\,,
\label{eq:power3}
\end{equation}
where we have explicitly written the powers of $\ell_{p}$, the Planck length. 
If we are interested in the effects of fluctuations within a finite
bandwidth, $(k,k +\Delta k)$, then we can write
\begin{equation}
\left\langle\left(\frac{\delta\rho}{\rho}\right)^{2}\right\rangle  =
\int d^{3}k\, e^{i\,\vec{k}\cdot\Delta\vec{x}}\,P_{k}(\eta_{s})
 \approx \Delta k\, k^2 \,  e^{i\,\vec{k}\cdot\Delta\vec{x}}\,P_{k}    \,.
\label{eq:power4}
\end{equation}
For the purpose of a rough estimate, let us take $\Delta k\approx k$
and $e^{i\,\vec{k}\cdot\Delta\vec{x}} \approx 1$. Then the corresponding
density perturbation is given by
\begin{equation}
\left(\frac{\delta\rho}{\rho}\right)_{rms}=
\sqrt{\left\langle\left(\frac{\delta\rho}{\rho}\right)^{2}\right\rangle}
\approx \sqrt{k^3\, P_{k}} \approx 
\ell_{p}^{2}\, |\eta_{0}|\, k^{3}\,  
\ln[a(\eta_s)]   \,.
\label{eq:power_estimate}
\end{equation}
Note that we have taken $a = 1$ at the end of inflation. As a result, 
$1/a(\eta_s)$ is the redshift factor between reheating and the last scattering
surface,
\begin{equation}
a(\eta_s) \approx \frac{E_R}{ {\rm 1 eV}} \,,
\end{equation}
where $E_R$ is the reheating energy scale.
We should have 
\begin{equation}
\left(\frac{\delta\rho}{\rho}\right)_{rms} \alt 10^{-4} \,,
\end{equation}
which leads to an upper bound on the duration of inflation
\begin{equation}
H\, |\eta_{0}| \alt 
10^{-4} \,\frac{H}{\ell_p^2\,k^3\, \ln\left(\frac{E_R}{ {\rm 1 eV}}\right)} \,.
   \label{eq:bound0}
\end{equation}
If $E_R$ is close to the scale of inflation, then the vacuum energy density
during inflation is $V_0 \approx E_R^4$, and
\begin{equation}
H^2 = \frac{8 \pi}{3}\, \ell_p^2\, V_0 \approx 
\frac{8 \pi}{3}\, \ell_p^2\,  E_R^4 \,. 
\end{equation}
Because we have chosen $a = 1$ at the end of inflation, the scale factor
today is
\begin{equation}
a_{\rm now} \approx 10^3 \,a(\eta_s) \approx 10^3\, \frac{E_R}{ {\rm 1 eV}}\,,
\end{equation}
and $k$ is related to the physical wavenumber today, $k_P$, by
$k = a_{\rm now}\, k_P$. 
Let $k_P = 2 \pi/\lambda$ correspond to the typical intergalactic separation
today, $\lambda \approx  2{\rm Mpc}$, or $k_P \approx 10^{-24} {\rm cm^{-1}}$.
Then we may combine the above relations to write the bound on the 
expansion factor during inflation as
\begin{equation}
H\, |\eta_{0}| \alt 10^{79}\, \left(\frac{10^{12} {\rm GeV}}{E_R}\right) \,,
   \label{eq:bound1}
\end{equation}
ignoring the weak logarithmic dependence upon $E_R$.

The $k^3$ dependence found in Eq.~(\ref{eq:power3}) indicates a non-scale
invariant spectrum of fluctuations which rises at shorter wavelengths.
The same dependence upon $k$ was found recently in a somewhat different 
context by Lombardo and Nacir~\cite{LN05}. (See Eq.~(68) of their paper.)
If observational data for $k_P > 10^{-24} {\rm cm^{-1}}$ were available,
then one might be able to obtain tighter bounds on the duration of inflation.
The smallest scales on which the cosmic microwave background has been observed
is about $5$ arcminutes~\cite{ACBAR}, which corresponds to 
$k_P \approx 10^{-24} {\rm cm^{-1}}$. Similarly, the role of higher values of
$k_P$ in large scale structure formation is unclear because of nonlinear
classical effects.

The effect we are considering depends upon transplanckian modes in the sense 
that the modes of the quantized electromagnetic or scalar fields which
give the dominant contribution have wavelengths much shorter than the Planck 
length. Let $L_i$ be a given proper length at the beginning of inflation, 
and let $L_f$ be the corresponding scale today stretched by the cosmological
expansion. These two scales are approximately related by
\begin{equation}
L_f = 10^3\, \left(\frac{E_R}{ {\rm 1 eV}}\right)\, H\, |\eta_{0}|\; L_i \,,
\end{equation}
as $H\, |\eta_{0}|$ is the expansion during inflation, ${E_R}/ {\rm 1 eV}$
is that between reheating and last scattering, and there has been an additional
expansion by a factor of about $10^3$ after last scattering. If we take
$L_f \approx  10^{24} {\rm cm}$, $E_R \approx 10^{12} {\rm GeV}$, and
$H\, |\eta_{0}| \approx 10^{79}$,  then $L_i \approx 10^{-46}\, \ell_p$.
It is well known that transplanckian modes play a crucial role in the
conventional approach to black hole evaporation~\cite{Hawking}. It is 
possible to obtain black hole evaporation without transplanckian modes,
but only at the price of introducing a  Lorentz non-invariant dispersion
relation~\cite{CJ96}.

\section{Second Model: Single Field Slow-Roll Inflation}
\label{sec:Second}

Let $\phi$ be an inflaton field which obeys the equation of motion
in a Robertson-Walker metric
\begin{equation}
\Box\phi= \frac{1}{a^{2}} \,\bigtriangledown^{2}\phi-
\frac{1}{a^{3}}\frac{\partial}{\partial t}
(a^{3}\frac{\partial\phi}{\partial t})=V'(\phi)\,,
\label{eq:inflaton}
\end{equation}
where $V(\phi)$ is a relatively flat potential, and $\bigtriangledown^{2}\phi$
is the flat space Laplacian operator. During a slow-roll phase, the second 
time derivative of $\phi$ is assumed to be small. If, in addition, the spatial
gradient terms are small, then 
\begin{equation}
3 \frac{\dot{a}}{a}\, \dot{\phi} \approx V'(\phi) \,.
\end{equation}
Note that $\theta_0 = 3 {\dot{a}}/{a} = 3H$ is the expansion of unperturbed 
deSitter spacetime. Now we wish to generalize this description to include
small spatial variations of the expansion. If the vorticity of the
comoving geodesics vanishes and the shear remains small, then 
the spacetime metric can be written as
\begin{equation}
ds^2 = -dt^2 +a^2(t,\mathbf{x}) (dx^2+dy^2+dz^2) \,.
\end{equation}
The expansion of the comoving geodesics, those  with four-velocity 
$u^\mu = \delta^\mu_t$ in these coordinates, is again
\begin{equation}
\theta = u^\mu_{;\mu} = 3 \frac{\dot{a}}{a} \, ,
\end{equation}
and the shear and vorticity vanish. The equation of motion for an inflaton
field with self-coupling $V(\phi)$ now becomes 
\begin{equation}
\Box\phi= \frac{1}{a^{3}} \,\nabla \cdot (a \nabla \phi) -
\frac{1}{a^{3}}\frac{\partial}{\partial t}
(a^{3}\frac{\partial\phi}{\partial t})=V'(\phi)\,.
\label{eq:inflaton2}
\end{equation}

Let each of the quantities $a$, $\phi$, and $\theta$ consist of a homogeneous
part and a small inhomogeneous perturbation:
\begin{equation}
a = a_0(t) + a_1(t,{\bf x}), \qquad
\phi = \phi_0(t) + \phi_1(t,{\bf x}), \quad {\rm and} \quad
\theta = \theta_0(t) + \theta_1(t,{\bf x})\,.
\end{equation}
We make the slow-roll approximation for the homogeneous part, $\phi_0(t)$,
which satisfies 
\begin{equation}
\dot{\phi_{0}}=-\frac{V'(\phi_{0})}{\theta_{0}}\, , \label{eq:inflaton3}
\end{equation}
where $\theta_{0}= 3 \dot{a}_0/a_0$. We retain the second time derivative of
$\phi_1(t,{\bf x})$, which satisfies
\begin{equation}
\ddot{\phi_1} +  \theta_0\, \dot{\phi}_1 + \theta_1\, \dot{\phi}_0 - 
\frac{1}{a_0^2}\,
\nabla^2 \phi_1 = - V''(\phi_0)\, \phi_1 \,, \label{eq:inflaton4}
\end{equation}
where we have expanded Eq.~(\ref{eq:inflaton2}) to first order in all of
the inhomogeneous perturbations and used
\begin{equation}
V'(\phi_0 + \phi_1) \approx V'(\phi_0) + V''(\phi_0)\, \phi_1 \,.
\end{equation} 
Let us consider the case where the potential is approximately linear
during the period of interest, so we may set $ V''(\phi_0) \approx 0$.
If we use Eq.~(\ref{eq:inflaton3}) and $\theta_0 = 3H$, 
we may write Eq.~(\ref{eq:inflaton4}) as
\begin{equation}
\ddot{\phi_1} + 3H \dot{\phi}_1 - \frac{1}{a_0^2}\, \nabla^2 \phi_1   = 
\frac{V'(\phi_{0})}{3 H}\; \theta_1 \,. \label{eq:inflaton5}
\end{equation}
Let us Fourier transform this equation and define
\begin{equation}
\delta \phi_k(t) = \int d^{3}k\, {\rm e}^{i\,\mathbf{k}\cdot\Delta\mathbf{x}}
\,  \phi_1(t,{\bf x}) \, ,
\end{equation}
and
\begin{equation}
\delta \theta_k(t) = \int d^{3}k\, 
{\rm e}^{i\,\mathbf{k}\cdot\Delta\mathbf{x}}\, \theta_1(t,{\bf x})\,.  
\end{equation}
We also change from comoving time $t$ to conformal time $\eta$.
Then $\delta \phi_k(\eta)$ satisfies
\begin{equation}
\frac{d^2 \delta \phi_k}{d \eta^2} + 
2 H a_0 \frac{d \delta \phi_k}{d \eta} + k^2 \delta \phi_k
= \frac{V'(\phi_{0})}{3 H}\,a_0^2\, \delta\theta_k \,. \label{eq:del_theta_k}
\end{equation}

Let $G(\eta,\eta')$ be a retarded Green's function for this equation which 
satisfies
\begin{equation}
\frac{d^2 G}{d \eta^2} + 2 H a_0 \frac{d G}{d \eta} + k^2  G
= \delta(\eta-\eta') \,,    \label{eq:G_eq}
\end{equation}
and $G(\eta,\eta')=0$ if $\eta < \eta'$.   It can be expressed as
\begin{equation}
G(\eta,\eta') = \frac{1}{W[\varphi_1(\eta')\, \varphi_2(\eta')]}\,
[\varphi_1(\eta_<)\,\varphi_2(\eta_>) 
\, - \,\varphi_1(\eta)\, \varphi_2(\eta')] \,.
\end{equation}
Here $\varphi_1$ and $\varphi_2$ are two linearly independent solutions
of Eq.~(\ref{eq:G_eq}) without the delta-function source term, and $W$ is
their Wronskian. Here $\eta_>$ and $\eta_<$ are, respectively, the greater
and lesser of  $\eta$ and $\eta'$. These solutions may be taken to be
\begin{equation}
\varphi_1(\eta) = \frac{1}{2} \sqrt{\pi} H |\eta|^\frac{3}{2}\,
H^{(1)}_\frac{3}{2}(\eta)
\end{equation} 
and
\begin{equation}
\varphi_2(\eta) = \frac{1}{2} \sqrt{\pi} H |\eta|^\frac{3}{2}\,
H^{(2)}_\frac{3}{2}(\eta)\,,
\end{equation} 
where $H^{(1)}_\frac{3}{2}$ and $H^{(2)}_\frac{3}{2}$ are Hankel functions.
The Wronskian becomes
\begin{equation}
W[\varphi_1(\eta')\, \varphi_2(\eta')] = \frac{i}{a_0^2(\eta')}\,.
\end{equation} 
If we set
\begin{equation}
F = \frac{V'(\phi_{0})}{3 H}\,a_0^2 \; \theta_1 \,,
\end{equation} 
then we can write the solution of Eq.~(\ref{eq:del_theta_k}) as
\begin{equation}
\delta \phi_k(\eta) = \int_{\eta_0}^0 d\eta' \, F(\eta')\, G(\eta,\eta')
=  \int_{\eta_0}^\eta d\eta' \, F(\eta')\, G(\eta,\eta')\,,
\end{equation} 
where we use the fact that $G(\eta,\eta')=0$ for $\eta'>\eta$.
In the expression for $G(\eta,\eta')$, we may now set $\eta_>=\eta$ and
$\eta_<=\eta'$. Let us now split the integration range into two parts as
\begin{equation}
\delta \phi_k(\eta) =  \int_{\eta_0}^{\eta_c} d\eta' \, 
F(\eta')\, G(\eta,\eta')
+  \int_{\eta_c}^\eta d\eta' \, F(\eta')\, G(\eta,\eta') \,.
                                                     \label{eq:split_int}
\end{equation} 
Here $\eta_c = -1/k$ is the horizon-crossing time for mode $k$ in conformal 
time. Note that in the first integral in the above expression, we have
$k|\eta| < 1 < k|\eta'|$, and in the second integral we have 
$k|\eta| < k|\eta'| < 1 $. If we assume that $k|\eta| \ll 1 \ll k|\eta'|$,
and use the limiting forms for the Hankel functions, we find
\begin{equation}
G \approx  \frac{\cos(k \eta')}{k^2 \eta'}\,.  \label{eq:G1}
\end{equation}
Similarly, if we assume $k|\eta| \ll k|\eta'| \ll 1 $, then we have
\begin{equation}
G \approx -\frac{1}{3} \, \eta' \,.        \label{eq:G2}
\end{equation}
Let us now make the approximation that we may use Eq.~(\ref{eq:G1}) in the
first integral in Eq.~(\ref{eq:split_int}), and Eq.~(\ref{eq:G2}) in the
second. 
The result is
\begin{equation}
\delta \phi_k(\eta) \approx -\frac{V'_0}{9 H^3} \left[3 \eta_c^2 
\int_{\eta_0}^{\eta_c} d\eta' \, \frac{\cos(\eta'/\eta_c)}{|\eta'|^3} 
- \int_{\eta_c}^\eta d\eta' \, \frac{1}{|\eta'|} \right] \, \delta \theta_k \,,
\end{equation}
where we have used $k=1/|\eta_c|$ and our assumption that $V'_0 =  V'(\phi_0)$
is approximately constant.

We now consider the effects of expansion fluctuations upon the evolution
of the inflaton field. The variance of $\delta \phi_k(\eta)$ is
\begin{eqnarray}
\langle(\Delta\phi_k)^{2}\rangle  &=&  \left(\frac{V_{0}'}{9H^{2}}\right)^{2}
\left[3(H\eta_c)^2 \int_{\eta_0}^{\eta_c} d\eta_1\, a^2_0(\eta_1) 
\cos(\eta_1/\eta_c) -  \int_{\eta_c}^\eta d\eta_1 \right] \nonumber \\
&\times& \left[3(H\eta_c)^2 \int_{\eta_0}^{\eta_c} d\eta_2\, a^2_0(\eta_2) 
\cos(\eta_2/\eta_c) -  \int_{\eta_c}^\eta d\eta_2 \right]  \nonumber \\
&\times& a_0(\eta_1) a_0(\eta_2) \; 
[\langle \delta\theta_k(\eta_1)\,\delta\theta_k(\eta_2)\rangle -
\langle \delta\theta_k(\eta_1)\rangle \langle\delta\theta_k(\eta_2)\rangle] \,.
\end{eqnarray}
The correlation function in the above integral is just the Fourier
transform of the coordinate space expansion correlation function given in
Eq.~(\ref{eq:theta_corr2}). It is convenient to evaluate the integrals in
coordinate space and to write variance of $\delta \phi(\eta)$ as
\begin{eqnarray}
\langle(\Delta\phi)^{2}\rangle & = & 
\left(\frac{8\pi V_{0}'}{9H^{2}}\right)^{2}
\left[3(H\eta_c)^2 \int_{\eta_0}^{\eta_c} d\eta_1\, a^2_0(\eta_1) 
\cos(\eta_1/\eta_c) -  \int_{\eta_c}^\eta d\eta_1 \right] \nonumber \\
&\times&  \left[3(H\eta_c)^2 \int_{\eta_0}^{\eta_c} d\eta_2\, a^2_0(\eta_2) 
\cos(\eta_2/\eta_c) -  \int_{\eta_c}^\eta d\eta_2 \right]  \nonumber \\
&\times& a^{-1}_0(\eta_1)\, a^{-1}_0(\eta_2)\, \; I(\eta_1,\eta_2) \, ,
\label{eq:Delphi}
\end{eqnarray}
where 
\begin{equation}
I(\eta_1,\eta_2) = \int_{\eta_{0}}^{\eta_{1}}d\eta\, a^{-1}_0(\eta)
\int_{\eta_{0}}^{\eta_{2}}d\eta'\, a^{-1}_0(\eta')\,{\cal E}(\Delta\eta,r) \,.
\label{eq:I}
\end{equation}

Here we have evaluated the variance of $\phi$ at the end of inflation,
$\eta=0$. This quantity is directly related to the density perturbations
in the post-inflationary era. This issue has been discussed extensively
by previous authors~\cite{MC81,GP82,Hawking82,Starobinsky82,BST83} 
in a context where $\Delta \phi$ arises from the intrinsic
quantum fluctuations of the inflaton field. However, the relation between
$\Delta \phi$ and the density perturbation is the same in the present context.
The basic idea is that spatial variations in $\phi$ cause different regions
to reheat at different times, with a typical time variation of
order $\Delta t \approx \Delta \phi/\dot{\phi_0}$. After reheating, the energy
density decreases as $\rho = \rho_R\, (t_R/t)^2$, where $\rho_R$ is the density
at reheating. This leads to density variations whose magnitude is of order
\begin{equation}
\frac{\delta\rho}{\rho} \approx \frac{2 \Delta t}{t_R} \approx 
 \frac{2 \Delta \phi}{t_R\, \dot{\phi_0}} 
\approx  \frac{6 H \,\Delta \phi}{t_R\,  V_{0}'} \,.
\end{equation}
In the last step, we have used Eq.~(\ref{eq:inflaton3}) with 
$ V_{0}' = V'(\phi_0)$. Thus we have
\begin{equation}
\left\langle\left(\frac{\delta\rho}{\rho}\right)^{2}\right\rangle=
\left( \frac{6 H }{t_R\,  V_{0}'}\right)^2 \;
\langle(\Delta\phi)^{2}\rangle \,,  \label{eq:del_rho_inf}
\end{equation}
with $\langle(\Delta\phi)^{2}\rangle$ given by Eq.~(\ref{eq:Delphi}).
The physical picture for the conversion of $\phi$-fluctuations into density
variations described above was first given by Guth and Pi~\cite{GP82}.
It is possible to give a more rigorous, gauge-invariant 
discussion~\cite{LL93,MBF92}.
However, for a single inflaton model, the results are essentially equivalent
to Eq.~(\ref{eq:del_rho_inf}). 
 
Consider first the case of an electromagnetic field. It is convenient
to express Eq.~(\ref{eq:EM_corr}) as
\begin{equation}
{\cal E}_{em} = 
-\frac{1}{480\pi^{4}}\left[\frac{d^{5}}{ds^{5}}
\frac{(r^{2}+3\Delta\eta^{2})^{2}}{(s-\Delta\eta^{2})}\right]_{s=r^{2}}\,,
\label{eq:EM_corr2}
\end{equation}
and to interchange the order of the $s$-differentiations and the 
integrations in Eq.~(\ref{eq:I}). In the limit of large $|\eta_{0}|$,
the result is
\begin{equation}
I \approx \frac{H^2 \eta_0^2}{40 \pi^4 r^6} \,,
\end{equation}
which is independent of $\eta_1$ and $\eta_2$ to leading order. Now the 
integrations on these variables in Eq.~(\ref{eq:Delphi}) can be performed
in terms of the cosine integral function, $ci(x)$, with the result 
\begin{equation}
\langle(\Delta\phi)^{2}\rangle   \approx
\frac{2\eta_{0}^{2}\,\eta_c^4\, (V_{0}')^{2}\,\ell_{p}^{4}}
{405\pi^{2}\, r^{6}}\, [1+6\, ci(1)]^2  \,,
\label{eq:Delphi2}
\end{equation}
where $ci(1) \approx 0.337$. Here we assume that $H |\eta_c| \gg 1$.
The corresponding expression for the conformal scalar field is smaller by
a factor of $1/3$.
In this model. the fluctuations grow at the same rate with increasing
$|\eta_{0}|$ as in the case of the model in Sect.~\ref{sec:First}.
This arises despite the fact that $\Delta\phi_k$ for an individual mode
only begins to grow after the mode leaves the horizon. The growth with
increasing $|\eta_{0}|$ comes from the fact that the expansion fluctuations
grow from the beginning of inflation, as described by 
Eq.~(\ref{eq:theta_corr2}).
If it were not for this growth, the factor of $\ell_{p}^{4}$ would tend
to make the effects of stress tensor fluctuations small compared to those
of the intrinsic quantum fluctuations of $\phi$.
The corresponding power spectrum of density perturbations is
\begin{equation}
P_{k} \approx
\frac{\ell_p^4\, H^2\,\eta_0^2 \,\eta_c^4 \,k^3} {540\,\pi^3 \, t_R^2}
\,  [1+6\, ci(1)]^2   \approx
\frac{\ell_p^4\, H^2\,\eta_0^2 }{540\,\pi^3 \, t_R^2 \, k}\, [1+6\, ci(1)]^2\,,
\label{eq:power5}
\end{equation}
where we set $|\eta_c| \approx 1/k$ only at the last step.
If we make the same estimates as were made in Eq.~(\ref{eq:power_estimate}),
we find
\begin{equation}
\left(\frac{\delta\rho}{\rho}\right)_{rms}
\approx \sqrt{k^3\, P_{k}} \approx 
10^{-2} \frac{\ell_{p}^2\, H|\eta_{0}|\, k} {t_{R}} \,.
\label{eq:power_estimate2}
\end{equation}
Note that both here and in the model of Sect.~\ref{sec:First}, we find 
a spectrum which is not scale invariant. Here
$({\delta\rho}/{\rho})_{rms} \propto k$.
If we again set $k_P \approx 10^{-24} {\rm cm^{-1}}$ and assume that 
$E_R \approx V_0^\frac{1}{4}$, we obtain the following upper bound on the 
duration of inflation:
\begin{equation}
H\, |\eta_{0}| \alt 
10^{45}\, \left(\frac{10^{12} {\rm GeV}}{E_R}\right)^3  \,.
   \label{eq:bound2}
\end{equation}
 This is considerably more restrictive than  Eq.~(\ref{eq:bound1}), but is 
still compatible with adequate inflation to solve the horizon problem.

\section{Summary and Discussion}
\label{sec:Final}

We have analyzed the effects of stress tensor fluctuations of conformally
invariant quantum fields in deSitter spacetime. One unexpected result
of this analysis is that expansion fluctuations grow during a deSitter
phase. This might be interpreted as due to the background spacetime altering
the anti-correlated fluctuations.
 In Minkowski spacetime, quantum fluctuations tend
to have strong correlations and anti-correlations. If one were to evaluate
the variance of the expansion in Minkowski spacetime, that is, compute
Eq.~(\ref{eq:theta_corr2}) with $a=1$, the result would be independent
of $\eta_0$ in the limit that $|\eta_0| \rightarrow \infty$. This is closely
related to the anticorrelations found in sampled energy density measurements
in flat spacetime. Similar anticorrelations are present when a charged
particle or a mirror is coupled to vacuum fluctuations in flat spacetime,
causing the mean squared velocity to approach a constant even in the absence
of dissipation~\cite{YF04,WL03,WL05}.
 It is the presence of nonconstant $a(\eta)$ functions
in the integrand of Eq.~(\ref{eq:theta_corr2}) which upsets the  flat 
spacetime cancellations and leads to a result which grows with increasing
$|\eta_0|$. 

It is this growth which allowed us to infer the constraints,
Eqs.~(\ref{eq:bound1}) and (\ref{eq:bound2}) on the duration of the
inflationary phase. These constraints may come as a surprize, as one usually
expects inflation to erase the memory of the past history of the universe.
It is certainly true that the exponential expansion quickly suppresses
classical perturbations. However, the effect discussed here amounts to
a type of quantum instability of deSitter spacetime: the cumulative effects
of passive metric fluctuations eventually lead to a spatially inhomogeneous
spacetime. The direct effect on the spacetime geometry grows rather slowly,
as reflected in the constraint Eq.~(\ref{eq:bound1}). However, when the
$\theta$-fluctuations couple to the inflaton field, the result can be a
stronger constraint, Eq.~(\ref{eq:bound2}). Both of these constraints
are consistent with adequate inflation to solve the horizon problem.
The possibility of effects which grow during inflation and react against
the expansion has been discussed by several authors as a possible solution 
for the cosmological constant problem. Among the effects considered
are the growth of long wavelength classical perturbations~\cite{GB02,BM04}
and backreaction due to quantum gravity effects~\cite{TW96}. The effect
discussed in the present paper is distinct from either of these effects,
especially as it produces an increasingly inhomogeneous universe rather 
than a backreaction against the cosmological constant.

Here is it worthwhile revisiting three of the assumptions which we made
in our analysis. One was the assumption that the $\delta \rho$ and
$\delta p$ terms in Eq.~(\ref{eq:theta1b}) are uncorrelated with the
term produced by stress tensor fluctuations. This assumption should hold
so long as the dominant source of density perturbations is other than the
stress tensor fluctuations. Of course, if inflation were to last sufficiently
long, this would no longer be true;  $\delta \rho$ and $\delta p$ would
be predominately due to these fluctuations. However, because the contributions
of stress tensor fluctuations are highly non-Gaussian and non-scale invariant,
they can give at best a very small contribution to the total primordial
fluctuation spectrum in our universe. Hence our assumption seems to
be justified. Another assumption which we have made is ignoring the shear
term in the Raychaudhuri equation. The comoving geodesics in deSitter
spacetime are certainly shear-free, so it is reasonable to start in a
state where $\sigma_{\mu\nu} = 0$. Shear could develop only if there were
large fluctuations of the Weyl tensor. This seems unlikely, but Weyl tensor
fluctuations from passive metric fluctuations need to be better understood. 
The third assumption is the sudden switching assumption first made in writing
Eq.~(\ref{eq:theta_corr}). This amounts to assuming that the expansion
fluctuations vanish before the $t=t_0$ hypersurface, and that the effects
of the stress tensor fluctuations appear suddenly after that time. In future
work, we plan to examine more general initial conditions to test the dependence
of our results upon the  initial conditions.

A particularly interesting possibility is that inflation lasted for
a time only slightly less than the constraints derived above. 
In this case, one 
would predict a small, but potentially observable effect from the quantum
stress tensor fluctuations. This effect is expected to manifest itself
in a non-Gaussian and non-scale invariant component in the density 
perturbations. It would also offer a possible probe of transplanckian physics.

\begin{acknowledgments}
We have benefitted from discussions with numerous colleagues. In particular,
L.H.F. would like to thank the participants of the 11th Peyresq workshop for
valuable comments. 
 This work was supported in part by the National Science Council of Taiwan
under Grant NSC95-2112-M-001-052-MY3,
and by the National Science Foundation under Grant PHY-0555754.
\end{acknowledgments}

\end{document}